\documentclass[]{spie}  %>>> use for US letter paper
\usepackage{amsmath,amsfonts,amssymb}
\usepackage{graphicx}
\usepackage[colorlinks=true, allcolors=blue]{hyperref}
\usepackage{bm}
\usepackage{float}
\usepackage{comment}

\title{How accurately can quantitative imaging methods be ranked without ground truth: An upper bound on no-gold-standard evaluation}

\author[a,b]{Yan Liu}
\author[a,b]{Abhinav K. Jha}
\affil[a]{Department of Biomedical Engineering, Washington University in St. Louis, St. Louis, MO, USA}
\affil[b]{Mallinckrodt Institute of Radiology, Washington University in St. Louis, St. Louis, MO, USA}

\authorinfo{Corresponding author: Abhinav K. Jha\\E-mail: a.jha@wustl.edu}

\begin{document} 
\maketitle

\begin{abstract}
Objective evaluation of quantitative imaging (QI) methods with patient data, while important, is typically hindered by the lack of gold standards. To address this challenge, no-gold-standard evaluation (NGSE) techniques have been proposed. These techniques have demonstrated efficacy in accurately ranking QI methods without access to gold standards. The development of NGSE methods has raised an important question: how accurately can QI methods be ranked without ground truth. To answer this question, we propose a Cramér–Rao bound (CRB)-based framework that quantifies the upper bound in ranking QI methods without any ground truth. We present the application of this framework in guiding the use of a well-known NGSE technique, namely the regression-without-truth (RWT) technique. Our results show the utility of this framework in quantifying the performance of this NGSE technique for different patient numbers. These results provide motivation towards studying other applications of this upper bound.
\end{abstract}


\section{INTRODUCTION}
\label{sec:intro}  % \label{} allows reference to this section

Quantitative imaging (QI), the extraction of quantifiable features from medical images, is being actively investigated to facilitate clinical decision-making.\cite{gatenby2013quantitative,rosenkrantz2015clinical} For example, the use of metabolic tumor volume (MTV) obtained from $^{18}$F-fluoro-deoxyglucose positron emission tomography (PET) to predict treatment response\cite{ohri2015pretreatment}, apparent diffusion coefficient measured using diffusion magnetic resonance imaging (MRI) to monitor cancer therapy response\cite{herneth2003apparent}, and radiotracer uptake in organ and tumor measured using quantitative single-photon emission computed tomography (SPECT) for dosimetry of radiopharmaceutical therapies\cite{li2022projection}. Multiple QI methods have been developed for each of these applications. For clinical translation of these methods, it is important to objectively evaluate these QI methods on the task of reliably measuring the underlying true quantitative value. Performing such evaluation with patient data is highly desirable but requires the knowledge of true quantitative value or a gold standard. Such gold standards are typically expensive, time-consuming, and in many cases, impossible to obtain. Thus, there is an important need for techniques to evaluate QI methods without a gold standard. 

To objectively evaluate QI methods in the absence of gold standards, no-gold-standard evaluation (NGSE) techniques have been proposed. A seminal work is the regression-without-truth (RWT) technique\cite{kupinski2002estimation,hoppin2002objective}. This technique posits a linear relationship between the measured and true values characterized by a slope, a bias, and an uncorrelated normally distributed noise term. Further, the technique assumes that the true values have been sampled from a bounded parametric distribution function with known bounds and unknown parameters. The RWT technique then estimates these linear relationship and true-distribution parameters using a maximum likelihood approach without access to the true quantitative values. A figure of merit (FoM) is then computed from the estimated parameters to rank the QI methods based on precision. The efficacy of this technique was demonstrated in comparing different software packages in cardiac SPECT for estimating cardiac ejection fraction\cite{kupinski2006comparing} and evaluating segmentation methods in cardiac sine MRI for estimating the left ventricular ejection fraction\cite{lebenberg2012nonsupervised}.

Subsequently, there have been multiple efforts to translate the RWT idea to broader clinical settings. The RWT technique assumes that the bounds of the true value distribution are known and that the noise between the different methods is independent, assumptions that may not hold in all clinical settings. Towards overcoming this issue, the technique was advanced to account for the cases where the bounds of the true value distribution are not known. The efficacy of the resultant technique was demonstrated in evaluating different SPECT reconstruction methods for estimating the mean activity uptake\cite{jha2016no} and PET segmentation methods on the task of measuring metabolic tumor volume\cite{jha2017practical}. The technique was then further advanced to model the correlated noise of the measurements given by different QI methods\cite{liu2022no}. The efficacy of this NGSE technique was demonstrated in evaluating SPECT reconstruction methods on the task of estimating mean activity uptake\cite{liu_no-gold-standard_2023}. Additionally, studies have been conducted to assess the sensitivity of the NGSE techniques to violations of certain other assumptions, as may occur in clinical settings. In these studies, it was observed that generally, the performance of the NGSE technique was relatively insensitive to violation of certain assumptions\cite{liu2023no}. 

With the NGSE techniques demonstrating efficacy in multiple settings, there is an important need for approaches to guide the use of these techniques. For example, assume that we want to use the NGSE technique to evaluate different PET lesion-segmentation methods on the task of estimating metabolic tumor volume (MTV). Consider that a dataset consisting of a finite number of patients has been collected. It is well known that the performance of NGSE techniques deteriorates as the number of patients that are input to the technique reduces\cite{kupinski2006comparing,jha2017practical}. Thus, guidance is needed on what is the best possible performance that can be achieved by the NGSE technique for the given patient dataset. This can provide guidance on whether the considered patient dataset is sufficient or if more patient data needs to be collected. Similarly, while NGSE techniques have demonstrated efficacy in ranking segmentation methods, their ranking accuracy may vary depending on the differences of the figure of merit for the considered segmentation methods. Here again, guidance may be needed on the best performance that can be achieved by the NGSE technique for evaluating the considered methods. 

The above needs lead to a central question, namely, how accurately can QI methods be ranked without ground truth. To answer this question, we developed a framework to quantify the best achievable ranking performance for no-gold-standard evaluation, i.e., the upper bound of the accuracy of NGSE techniques on correctly ranking QI methods. The proposed framework was developed for NGSE techniques that are based on the premise that there is a statistical relationship between the true quantitative value and the quantitative values measured using the different QI methods, and where the parameters of this relationship can be estimated without any knowledge of the true quantitative value. The RWT technique and other recently proposed NGSE techniques belong to this class. We first provide a brief description of the theory of this upper bound.  

\section{Methods}
\subsection{Theory}
\label{sec:theory}
Consider a set of $K$ QI methods that are used to measure certain quantitative values from a patient population consisting of $P$ patients. Denote the true quantitative value for $p^\mathrm{th}$ patient as $a_p$ and the corresponding measurement given by $k^\mathrm{th}$ QI method as $\hat{a}_{p,k}$. Existing NGSE techniques assume a linear relationship between the measured and true values characterized by a slope, bias and zero-mean Gaussian distributed noise term. For $k^\mathrm{th}$ QI method, denote the slope and bias as $u_k$ and $v_k$, respectively. The noise term is denoted as $\mathcal{N}\left(0,\bm{\Sigma}\right)$, where $\bm{\Sigma}$ denotes the covariance matrix. For $p^\mathrm{th}$ patient, the relationship between the measured quantitative values and the true value can be written as:
\begin{equation}\label{eq1}
    \begin{bmatrix}
        \hat{a}_{p,1}\\\hat{a}_{p,2}\\ \vdots\\\hat{a}_{p,K} 
    \end{bmatrix} = 
    \begin{bmatrix}
        u_1a_p+v_1\\
        u_2a_p+v_2\\ \vdots
        \\ u_ka_p+v_k
    \end{bmatrix}
    +\mathcal{N}\left(0,\bm{\Sigma}\right).
\end{equation}

Denote the vector $\left[u_1,v_1,\ ,u_2,v_2\ ...,\ u_K,v_K\right]$ by $\bm{\Theta}$ and the measurements $\left[\hat{a}_{p,1},\ \hat{a}_{p,2},\ ...,\ \hat{a}_{p,K}\right]$ by $\hat{\bm{A}_p}$. Based on Eq.\ref{eq1}, we can obtain the probability distribution of $\hat{\bm{A}}_p$, but that depends on the true value. Thus, to estimate these parameters, the true values are needed. To address this issue, the NGSE techniques assume that the true quantitative values are sampled from a parametric distribution $\text{pr}\left(a_p|\bm{\Omega}\right)$ characterized by a parameter vector $\bm{\Omega}$. Denote all the parameters $\left\{\bm{\Theta},\bm{\Sigma},\bm{\Omega}\right\}$ as $\bm{\Gamma}$. By further assuming that the true values of different patients are independent, the log-likelihood function of all the measurements can be derived as follows:
\begin{equation}\label{eq2}
    \Lambda\left(\bm{\Gamma}\big|\left\{\hat{\bm{A}}_p\right\}\right) = \sum_{p=1}^P\ln \int \text{pr} \left(\hat{\bm{A}}_p|a_p,\bm{\Theta},\bm{\Sigma}\right)\text{pr}\left(a_p|\bm{\Omega}\right) da_p.
\end{equation}
This log-likelihood expression provides the basis for developing the CRB-based framework. More specifically, based on Eq.\ref{eq2}, we can obtain the Fisher information matrix of $\bm{\Gamma}$. By taking the inverse of this matrix, we obtain the CRB of $\bm{\Gamma}$, denoted as $\text{CRB}\left(\bm{\Gamma}\right)$. 

In NGSE techniques, the FoMs used to rank the QI methods are computed from the estimated parameters $\bm{\Gamma}$. Most commonly, for NGSE techniques that assume a linear relationship between the true and measured values, the ratio of the noise standard deviation and slope term, referred to as the noise-to-slope ratio (NSR), is used to rank the QI methods based on precision. Denote the FoM for $k^\mathrm{th}$ QI method as $\gamma_k$ and let $\bm{\gamma}=\left[\gamma_1,\ \gamma_2,\ ...,\ \gamma_K\right]$. Denote the Jacobian matrix of the function that computes $\bm{\gamma}$ from $\bm{\Gamma}$  as $\bm{J}$. The CRB of $\bm{\gamma}$ can be computed using following equation:
\begin{equation}
    \text{CRB}\left(\bm{\gamma}\right) = \bm{J}\left[\text{CRB}\left(\bm{\Gamma}\right)\right]\bm{J}^T.
\end{equation}
For an unbiased estimator, the best performance for estimating the parameters is that the variance of the estimated value achieves the CRB, in which case the distribution of the estimated parameters can be considered approximately normal distributed\cite{barrett2013foundations}. In this case, the distribution of the estimated FoMs $\hat{\bm{\gamma}}$ is given by:
\begin{equation}\label{eq4}
    \hat{\bm{\gamma}}\sim \mathcal{N}\left(\bm{\gamma},\text{CRB}\left(\bm{\gamma}\right)\right).
\end{equation}

As the distribution of the estimated FoMs can be obtained when the unbiased estimator achieves the best performance, an upper bound of ranking the QI methods can thus be computed. For example, if three QI methods are being evaluated and the true FoM values have the order $\gamma_1<\gamma_2<\gamma_3$. The upper bound for correctly ranking the QI methods, denoted as $U_\mathrm{cr}$ can be computed using following equation:
\begin{equation}
    U_\mathrm{cr} = \iiint_{\hat{\gamma}_1<\hat{\gamma}_2<\hat{\gamma}_3} \text{pr}\left(\hat{\gamma}_1,\hat{\gamma}_2,\hat{\gamma}_3\right)d^3\hat{\bm{\gamma}}.
\end{equation}
A similar upper bound can be computed for correctly identifying the best QI method.

\subsection{Demonstrating utility of the proposed framework}

We applied the framework to guide the use of the RWT technique for evaluation of QI methods without gold standards. Note that the RWT technique assumes a linear relationship between the measured quantitative values and the true values, and the noise yielded by different QI methods is uncorrelated. The technique then uses estimated noise-to-slope ratio to rank the different QI methods on the task of precisely estimating the underlying quantitative value\cite{hoppin2002objective}.

We considered three hypothetical QI methods, in each of which, there was a linear relationship between the true and measured values. More specifically, the values of slope were set to $\left\{0.6,0.7,0.8\right\}$ for each of the three methods. Similarly, the values of bias were set to $\left\{-0.1,0,0.1\right\}$ and the values of noise standard deviation were set to $\left\{0.03,0.05,0.08\right\}$ for the three QI methods. The true values were sampled from a beta distribution. The parameters for beta distribution, $\left\{\alpha,\beta\right\}$, were set to $\left\{1.5,2\right\}$. We used the proposed framework to compute the upper bound for correctly ranking the three hypothetical QI methods as the number of patient samples that was input to the RWT technique was varied. The number of patients was varied from 5 to 400. 

For comparison, numerical studies were conducted to compute the experimental ranking performance of the RWT technique. For each considered number of patients, we sampled true values from the beta distribution. From these true values, we generated synthetic measurements for three hypothetical QI methods. These measured values were linearly related to the true values by the above slope, bias, and Gaussian noise term. The measurements were then input into the RWT technique to obtain the estimates of the linear relationship parameters $\left\{\bm{\Theta},\bm{\Sigma}\right\}$. The FoM, which is the noise-to-slope ratio, was then computed to rank the QI methods based on precision. We repeated the experiment for 100 noise realizations to compute the accuracy of the RWT technique in correctly ranking the three QI methods. We refer to these results as experimentally derived in Section \ref{sec:Results}. 

\section{Results}
\label{sec:Results}
\begin{figure}[H]
    \centering
    \includegraphics[scale=0.7]{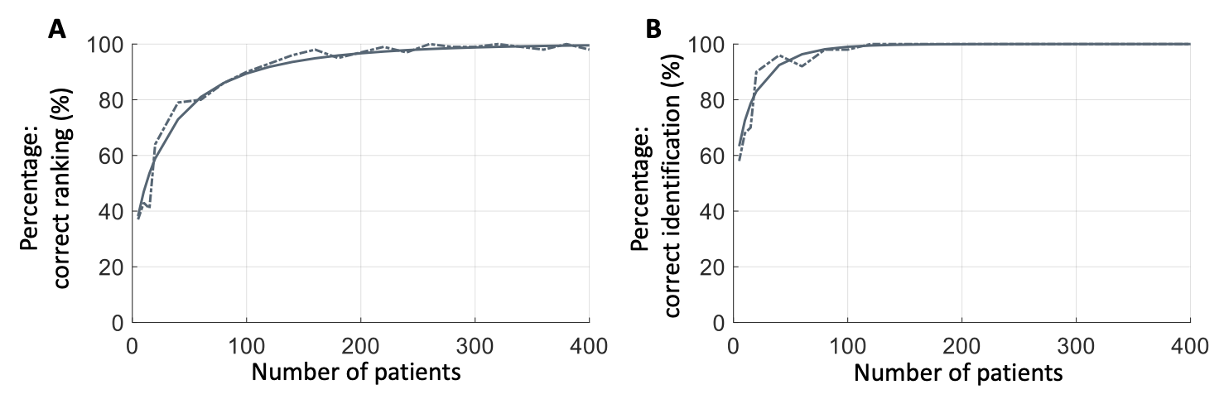}
    \vspace{0.2cm}
    \caption{A comparison of the CRB-derived upper bound for no-gold-standard evaluation (solid lines) vs. the experimentally derived accuracy of the RWT technique (dashed line) in (A) correctly ranking all the QI methods and (B) correctly identifying the most precise method with different number of patients as inputs to the technique.}
    \label{fig:1}
\end{figure}

Fig.\ref{fig:1} presents the comparison of the upper bounds with the experimentally derived accuracy of the RWT technique in correctly ranking the QI methods and identifying the QI method that yields the most precise measurements. We observe a good match between the framework-predicted upper bound and the experimentally derived performance of the RWT technique. Further, we observe that the use of these upper bounds can predict the number of patients required to achieve a specific accuracy in ranking the methods (Fig. \ref{fig:1}A) and in correctly identifying the most precise method (Fig. \ref{fig:1}B). For example, if given a dataset of 300 patients, the best possible performance yielded by the RWT technique can be an accuracy of 100\% in ranking this set of QI methods. However, if only 100 patients were available, the RWT technique can yield accurate ranking of the methods only in 90\% cases at best. If the goal is just to identify the most precise method without necessarily ranking the methods, then with this 100-patient dataset, the RWT technique can, at its best, be accurate 100\% of the times. This shows the application of the proposed framework in determining the number of patients required to achieve a certain accuracy when using the NGSE technique.

\section{Discussion and Conclusion}
To answer the question of how accurately can QI methods be ranked without gold standards, we proposed a Cramér–Rao bound-based framework. This framework quantifies the upper bound of correctly ranking the QI methods in the absence of gold standards.  We then presented an application of the framework in guiding the use of the RWT technique. The results demonstrated the utility of this framework in quantifying the upper bound of the performance of no-gold-standard evaluation for different number of patients input to the technique.

This upper bound provides a principled approach to guide the use of NGSE techniques. For a given set of QI methods, the upper bound can provide the best achievable ranking performance for a given number of patients. If this upper bound is low, caution is advised when using the NGSE technique with patient data. Another potential application of the framework is to assess the performance of NGSE techniques in accurately ranking different QI methods that may have varying degrees of separation of the figure of merit. This application of the upper bound needs further investigation.

The upper bound can also be used to compare the performance of evaluating QI methods with and without the gold standard. This helps to determine if obtaining the gold standard is necessary for evaluating the QI methods. Considering the example, as presented in the introduction, of evaluating segmentation methods on the task of estimating metabolic tumor volume. For this example, a potential gold standard could be provided by the resected tumor. However, obtaining such gold standards for a patient population can be a time-consuming and resource-intensive task. Further, even with available gold standards, statistical estimation techniques such as least squares still need to be used to estimate the FoMs, and thus these estimated FoMs may have their own inaccuracies\cite{obuchowski2015quantitative}. Hence, a quantitative assessment of the difference between the ranking performance of no-gold-standard evaluation and the evaluation with gold standards can help decide whether gold standards are necessary. If the difference between these two cases is acceptable, no-gold-standard evaluation can circumvent the tedious and time-consuming process of obtaining gold standards. Investigating this application of the proposed framework is another area of future research.  

One limitation of the proposed framework is that it assumes that the NGSE technique yields unbiased estimates of the figures of merit. However, NGSE techniques could potentially yield biased estimates. For example, to address the issue of the NGSE techniques requiring a large number of patient samples, a maximum-a-posteriori-based NGSE technique was proposed\cite{jha2015incorporating}. Such an estimator can be biased. Deriving bounds for such biased NGSE techniques is an area of future investigation. Next, although the assumptions of the NGSE technique are reasonable, they could be potentially violated in clinical settings\cite{jha2016no,liu2023no}. For example, while the linear relationship of the measured and true quantitative values is desirable in QI, this assumption may not always hold true in clinical scenarios. Thus, it is important to assess the impact of the violation of these assumptions on the upper bound. 

In conclusion, we developed a framework to quantify the upper bound in correctly ranking the methods without gold standards. This framework can also quantify the upper bound for correctly identifying the best QI method. We observed the utility of this upper bound in assessing the number of patients required by a specific NGSE technique, namely the RWT technique, to obtain a certain degree of accuracy. Our results motivate further investigations of the applications of the proposed framework.

\acknowledgments % equivalent to \section*{ACKNOWLEDGMENTS}       
This work was supported by the National Institute of Biomedical Imaging and Bioengineering of the National Institute of Health under grants R01-EB031051, R01-EB031051-S1 and R01-EB031962. We would like to thank Barry A. Siegel, MD, for inputs related to the clinical context of this work, and Zekun Li for helpful discussion.

% References
\bibliography{report} % bibliography data in report.bib
\bibliographystyle{spiebib} % makes bibtex use spiebib.bst

\end{document}